\documentclass[journal]{IEEEtran}
\usepackage{cite}
\usepackage{ulem}
\usepackage{bm}
\usepackage{graphicx}
\usepackage{amsmath,bm}
\usepackage{array}
\usepackage{mdwmath}
\usepackage{amssymb}
\usepackage{mdwtab}
\usepackage{stfloats}
\usepackage{amsmath,amsthm}
\usepackage{threeparttable}
\usepackage{color}
\usepackage{ulem}
\usepackage[ruled,linesnumbered]{algorithm2e}
\usepackage{multirow}
\usepackage[caption=false,font=normalsize]{subfig}
\usepackage{fancyhdr}

\makeatletter
\newcolumntype{"}{@{\hskip\tabcolsep\vrule width 3pt\hskip\tabcolsep}}
\makeatother

\begin{document}
	
	\title{An Efficient Pre-Processing Method for 6G Dynamic Ray-Tracing Channel Modeling}
	\author{Songjiang Yang, \IEEEmembership{Member,~IEEE}, Cheng-Xiang Wang, \IEEEmembership{Fellow,~IEEE}, Yinghua Wang,\\  Jie Huang, \IEEEmembership{Member,~IEEE}, Yuyang Zhou, and el-Hadi M. Aggoune, \IEEEmembership{Life Senior Member,~IEEE}
	\thanks{This work was supported by the National Natural Science Foundation of China (NSFC) under Grants 61960206006, 62401644 and 62271147, the Fundamental Research Funds for the Central Universities under Grant 2242022k60006, the Key Technologies R\&D Program of Jiangsu (Prospective and Key Technologies for Industry) under Grants BE2022067, and BE2022067-4, the Start-up Research Fund of Southeast University under Grant RF1028623029, the Research Fund of National Mobile Communications Research Laboratory, Southeast University, under Grant 2024A05, and AI and Sensing Technologies Research Center, University of Tabuk, KSA under Grant 1445-200. (\textit{Corresponding Author: Cheng-Xiang Wang and Yinghua Wang})}

	\thanks{S. Yang and Y. Wang are with the Pervasive Communication Research Center, Purple Mountain Laboratories, Nanjing, 211111, China, and also with the  National Mobile Communications Research Laboratory, School of Information Science and Engineering, Southeast University, Nanjing 211189, China (e-mail: \{yangsongjiang, wangyinghua\}@pmlabs.com.cn).
		
	C.-X. Wang, J. Huang, and Y. Zhou are with the National Mobile Communications Research Laboratory, School of Information Science and Engineering, Southeast University, Nanjing 211189, China, and also with the Pervasive Communication Research Center, Purple Mountain Laboratories, Nanjing 211111, China (e-mail: \{chxwang, j\_huang, yuyzhou\}@seu.edu.cn). 
	
	E. M. Aggoune is with AI and Sensing Technologies Research Center, University of Tabuk, Tabuk 47315/4031, Saudi Arabia (e-mail: haggoune@ut.edu.sa).}	
	\thanks{Copyright (c) 2024 IEEE. Personal use of this material is permitted. However, permission to use this material for any other purposes must be obtained from the IEEE by sending a request to pubs-permissions@ieee.org.}
	}
	\markboth{IEEE Transactions on Vehicular Technology,~Vol.~XX, No.~XX, Month~2024}
	{Shell \MakeLowercase{\textit{et al.}}: Bare Demo of IEEEtran.cls for Journals}
	\maketitle
	
	\begin{abstract}	
	The ray-tracing is often employed in urban areas for channel modeling with high accuracy but encounters a substantial computational complexity for high mobility scenarios.
	In this paper, we propose a novel pre-processing method for dynamic ray-tracing to reduce the computational burden in high-mobility scenarios by prepending the intersection judgment to the pre-processing stage.
	The proposed method generates an inter-visibility matrix that establishes visibility relationships among static objects in the environment considering the intersection judgment.
	Moreover, the inter-visibility matrix can be employed to create the inter-visibility table for mobile transmitters and receivers, which can improve the efficiency of constructing an image tree for the three-dimensional (3D) dynamic ray-tracing method.
	The results show that the proposed pre-processing method in dynamic ray-tracing has considerable time-saving compared with the traditional method while maintaining the same accuracy. 
	The channel characteristics computed by the proposed method can well match to the channel measurements.
	\end{abstract}
	
	\begin{IEEEkeywords}
	 Channel model, dynamic ray-tracing, high mobility scenarios, Image method, mobile communications
	\end{IEEEkeywords}
	
	\section{Introduction}
	\normalem
	The sixth-generation (6G) communications enable a new range of applications and global coverage scenarios, e.g., vehicular communications, unmanned aerial vehicle (UAV) communications, and satellite communications \cite{CXWang2020VTM}.
	Moreover, the 6G communication system is expected to provide higher data rate, lower latency, and more accurate localization compared with the fifth-generation (5G) communication system \cite{CXWang2023COMST}.
	The channel model research plays a crucial role in the design and performance evaluation of wireless communication systems to realize the emerging applications \cite{Molisch2011}.
	To support these 6G applications, a highly accurate channel model should be considered in these scenarios.
	
	Future 6G vehicular or UAV communications require the propagation channel prediction for the mobile transmitter and receiver with high dynamic mobility \cite{SYANG2022TVT, Ma2023IoTJ, Zhu2022TVT}.
	For mobile transmitter and receiver scenarios, the geometry-based stochastic model (GBSM) and beam-domain channel model (BDCM) are widely used to describe the wireless channel characteristics with low computational burden and good accuracy in general cases \cite{Chang2023TWC, Hou2023TNSE}.
	However, since the GBSM and BDCM use a set of general environment parameters to describe the simulation geometric environment, the accuracy for some specific propagation environment cases may not achieve the 6G accuracy requirement of channel model \cite{CXwang2023Tcom, CXWang2022TVT}.
	
	From channel modeling accuracy aspects, the deterministic ray-tracing method is commonly used for propagation channel modeling with high accuracy, which can achieve the 6G accuracy requirement of channel model in specific propagation environments \cite{He2019TUT, Yang2020RS}. 
	The ray-tracing method can find all the probable pathways for the signal traveling from the transmitter to the receiver by using geometrical optics (GO).
	Then, the ray-tracing method computes the electromagnetic field by Maxwell's law and uniform theory of diffraction (UTD) methods.
	The ray-tracing method can identify the dominant rays by considering the building geometry present in the environment between the transmitter and receiver \cite{Yun2015ACCESS, Qiu2022WCL}.
	Hence, the computational complexity of ray-tracing with detailed environment information is too high to simulate the mobile transmitter and/or receiver case.
	How to model the environments, e.g., trees and street lamps, with low complexity and high accuracy in channel prediction is always a main topic in the ray-tracing study \cite{Vitucci2019TAP, Lee2019AWPL}.
	
	Another consideration for improving ray-tracing is the development of the ray-tracing algorithm to reduce the computation time, which may help the ray-tracing method to support 6G communication channels with high mobility scenarios \cite{SYang2019ISAP}.
	There has been a lot of recent interest in reducing complexity and improving efficiency of the ray-tracing algorithm.
	In \cite{Wang2024VTC}, the authors proposed an improved bounding volume hierarchy (BVH) considering static and dynamic objects in the vehicle-to-vehicle (V2V) dynamic scenario to enhance the computational efficiency.
	In \cite{Tiberi2009TAP}, the authors used a parallel ray approximation technique to reduce the complexity of  ray-tracing by assuming a constant number of rays in a wavelength region.
	However, it may not be suitable for more complex scenarios or applications requiring real-time performance for mobile transmitters and receivers.
	In \cite{Suga2021WCL}, the authors proposed a novel ray-tracing acceleration method based on total variation norm minimization (TVNM) for radio map interpolation.  
	In TVNM algorithm, the edge between line-of-sight (LoS) and non-LoS (NLoS) regions was recognized to improve the accuracy of the interpolation, but the proposed algorithm can only predict the path loss information. 
	In \cite{Bilibashi2020EuCap, Bilibashi2023TAP}, the authors introduced a new dynamic ray-tracing model that can predict propagation channels through a single run using a dynamic environment database based on the extrapolation concept.
	The proposed dynamic ray-tracing algorithm was effective within the specified coherence time range and was suitable for two-dimensional (2D) scenarios.
	Nonetheless, the procedures for ray creation and suppression, and the generation of coherence time were not addressed.
	In a stochastic channel description approach, the coherence time equivalent is determined by the stationary interval as defined in \cite{Gan2018TVT}, where the delay or angular power profiles of the channel can be seen as stable. 
	This refers to a specific time or space interval during which the channel can be considered as wide sense stationary.
	The dynamic ray-tracing model was further developed using the numerical approach to track interaction points in the ray-tracing process for V2V communications. 
	Furthermore, the ray creation and suppression process and the rule of extrapolation time were deeply studied \cite{Quatresooz2021TVT}.
	The creation and suppression of rays in coherence time for dynamic ray-tracing have not been explored in the literature for high mobility scenarios. 
	
	One of the ray creation and suppression methods in the ray-tracing is the visibility checking of each face to support generating the image tree \cite{Geok2018IJCAP}.
	The application of efficient computer graphics algorithms has proven successful in identifying visible 2D/ three-dimensional (3D) surfaces for fixed transmitters.
	However, checking the visibility independently for each transmitter position and achieving accurate results can be time-consuming during pre-processing \cite{Agelet2000TVT, Ng2007TVT}. 
	Accordingly, the ray-tracing method in high mobility scenarios is currently restricted due to its limited applicability for longer pre-processing times.
	To reduce the pre-processing time, an efficient ray-tracing algorithm relying on a pre-computed database of intra-visibility of walls and edges was proposed, which is suitable for a mobile transmitter operating in urban environments \cite{Hussain2019TAP}.
	Furthermore, an efficient ray-tracing algorithm using a pre-computed visibility table to calculate the direct visibility of walls and edges for a mobile receiver was proposed.
	The visibility relationship was then used to accelerate the ray validation process by reducing the image tree in vehicular communications \cite{Hussain2022TAP}.
	However, this pre-computed intra-visibility table and visibility table are suitable for the 2D transmitter and receiver straight line moving, which cannot support the high mobility scenario in 6G.
	We note that how to reduce the pre-processing time for the ray-tracing method with the 3D trajectories of transmitter and receiver moving has not been sufficiently studied.
	 		 
	In this paper, we propose an efficient 3D inter-visibility pre-processing framework to accelerate ray-tracing computations and compute channel characteristics.
	More specifically, we propose the 3D inter-visibility matrix to list all visible faces and edges for all static scatterers in high mobility scenarios, where the interface relations information issued to build the 3D angular bouncing ranges.
	Additionally, we also propose a method to generate the 3D inter-visibility table for high mobility scenarios to find the ray creation or suppression points for generating the coherence time for different segments of transmitter or receiver trajectory.  
	The main contributions and novelties of this paper are summarized as follows: 
	\begin{itemize}
	\item 
	An efficient 3D pre-processing framework for dynamic ray-tracing is proposed to speed up the ray-tracing simulation time for high mobility scenarios. 
	This framework contains the following two parts: capturing the static environmental status to generate a 3D inter-visibility matrix and incorporating the mobile status of transmitters and receivers to build a 3D inter-visibility table. 
	In the 3D inter-visibility matrix, the intersection judgments of each static scatterer's face are prepended into the pre-processing stage. 
	Moreover, the 3D inter-visibility table shows the segment for the mobile trajectories to generate the coherence time of simulation situations, which is the key parameter to operate the dynamic ray-tracing method. 
	\item 
	The 3D inter-visibility matrix of the static building in high mobility scenarios is proposed by using the interface relation information and angular bouncing range to catalog visible faces and edges. 
	The matrix lists all visible faces and edges for each face and edge of all buildings in the environment, where first-order and multiple-order reflection information for these faces can be computed based on this information. 
	The image tree of all buildings is generated and stored in the inter-visibility matrix for repeated utilization.
	\item 
	The 3D inter-visibility table is proposed based on the 3D inter-visibility matrix and the trajectories to minimize the number of ray checks and ray trace procedures in high mobility scenarios. 
	The visibility judgment between the moving transmitter/receiver and the static scatterers based on the 3D inter-visibility matrix is proposed to find the ray creation or suppression point. 
	Moreover, the coherence time along the trajectory is found at the points of ray creation or suppression based on the 3D inter-visibility table to support dynamic ray-tracing.
	\end{itemize}  

	The rest of this paper is organized as follows. 
	In Section~II, the ray-tracing pre-processing method of inter-visibility matrix for static scenarios is presented. 
	In Section III, the ray-tracing pre-processing method of inter-visibility table for mobile scenarios is presented.
	In Section IV, results about the accuracy and time-saving of the proposed algorithm are shown. 
	Finally, in Section V, our main conclusions are drawn.
	\section{3D Inter-visibility Matrix for Static Status }
	The most time-consuming procedure for a ray-tracing algorithm using the Image method is the step to generate the image tree of the scenarios.
	The visible faces and edges that satisfy the maximum order of the reflection or diffraction ray can be identified and recorded to form the image tree, where the visibility of each face or edge in buildings would be checked recursively.
	Moreover, the visibility between faces and faces or edges, and visibility between faces and transmitter or receiver are used to build the inter-visibility list, which can determine possible propagation paths between transmitter and receiver. 
	To reduce the creation and traversal step of the image tree, the pre-processing method by generating an inter-visibility matrix of simulation scenarios can be executed to speed up the ray-tracing algorithm \cite{Saeidi2012TAP}.	

		\begin{figure*}[t]
			\centering
		\subfloat[3D view]{\includegraphics[width=0.33\linewidth]{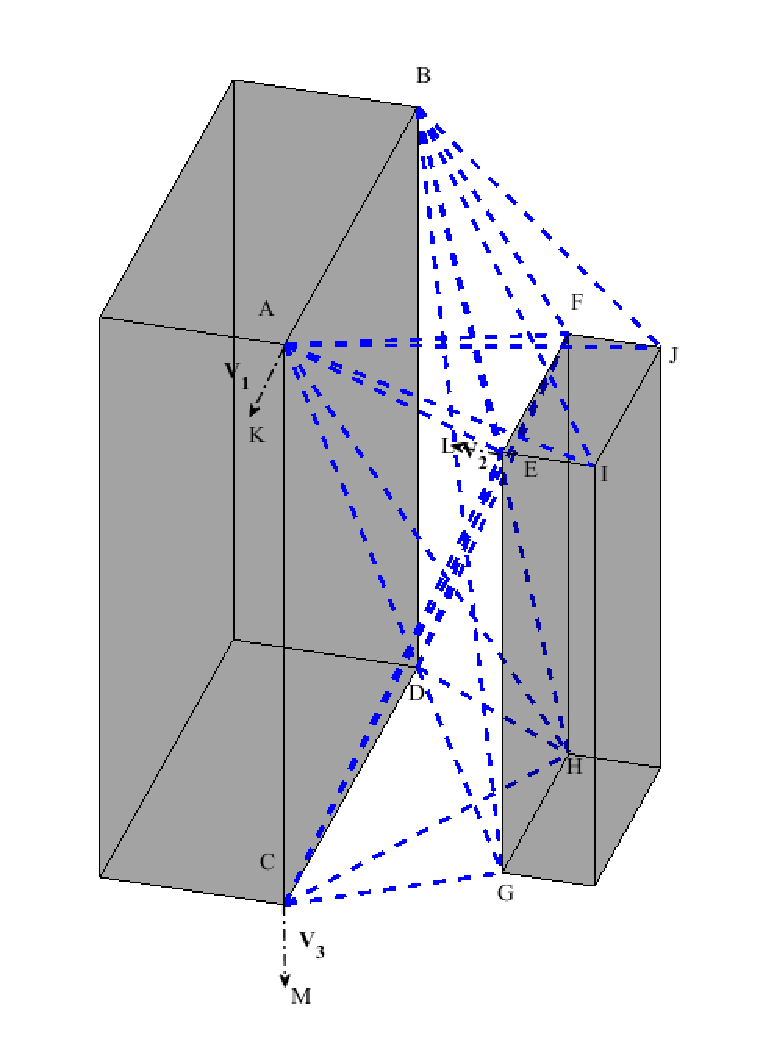}}
		\label{3D}
			\hfil
			\subfloat[xy-plane]{\includegraphics[width=0.33\linewidth]{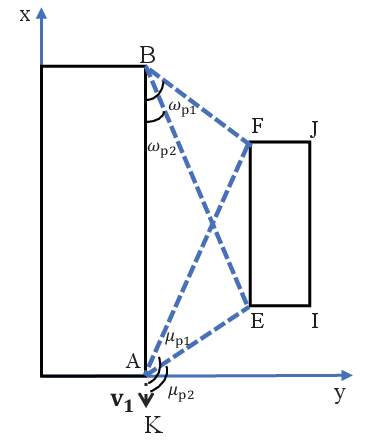}}
			\hfil
				\subfloat[yz-plane]{\includegraphics[width=0.33\linewidth]{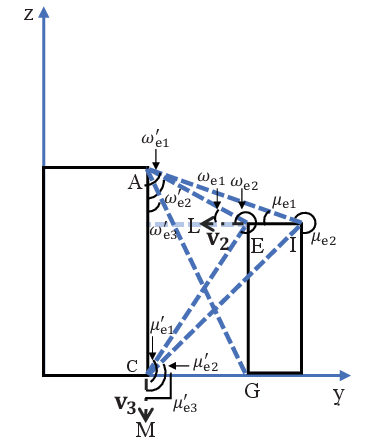}}
			\hfil
			\caption{Determination visibility angles of a visible face in (a) 3D view, (b) xy-plane view, and (c) yz-plane view at first-order reflection. The blue dash lines express the angular bouncing range.}
				\label{fig:sysmodelANGover180}
			\end{figure*}	
	\subsection{3D Inter-visibility Matrix}
	The 3D inter-visibility matrix can be used to generate the image tree of all objects and find the available path.
	The 3D inter-visibility matrix stores the index of visibility vertex points for the building, angular bouncing range, and interface relations in practice.
	The 3D inter-visibility matrix can be used to create the first-order image point of the faces and edges for all possible transmitter and receiver positions, which can be used to develop the second-order or higher-order reflection ray. 
	The image tree of the static building is stored in a chain structure and can be associated with the face ID and the interface relation information in the scenario database.
 
	In a 3D static scenario, the 3D angular information for each face and edge can describe the relationship between each building.
	However, generating 3D angular information directly for a 3D inter-visibility matrix is complex and challenging \cite{Cuinas2020APM}. 
	To overcome this problem, 3D scenarios can be divided into two 2D scenarios to compute 2D angular information, allowing the 3D inter-visibility matrix to be composed of two 2D inter-visibility matrices (xy-plane and xz-plane or yz-plane), as shown in Fig. \ref{fig:sysmodelANGover180}.
	In our case, the same position points in the two 2D planes are related to each other to generate the full 3D scenario. 
	For static building scenario computation, it can be simplified by considering a 2D horizontal plane and a height, i.e., a 2.5D concept, because there is no rotation or other 3D motion in the static scenario. 
	However, if the trajectory of mobile transmitter and receiver with height changing is considered, they have the 3D motion leading to the 3D inter-visibility matrix. 
	The interface relation in the inter-visibility matrix labels the xy-plane, yz-plane or geometry relationship of each face. 
	The angular bouncing range in the inter-visibility matrix records the angular information for each face and edge.

	\subsection{Single Reflection Procedure for Static Status}
	After introducing the inter-visibility matrix, the single reflection case for these static buildings is initially introduced.
	Since buildings in urban scenarios are static, the visibility between buildings remain unchanged and can be represented using constant angular information.
	Moreover, the visibility area between an edge and a face can be defined by two angular information, where an edge from one face is linked to two edges on another face. 
	The single reflection is from the transmitter to the aim face through the reference face, where the reference face has the reflection point and the aim face received the reflected ray.
	In generating the inter-visibility matrix step, the transmitter position is unknown and all possible visibility between the static buildings should be contained. 
	Accordingly, the visibility between the transmitter and the reference face is assumed all visible to keep all possible between the static buildings. 
	For reference face and aim face in single reflection, the two faces are close to each other and no other obstacles for the single reflection. 
		
	In Fig. \ref{fig:sysmodelANGover180}, two buildings are illustrated in 3D view and two 2D views with xy-plane and yz-plane as examples to help understand the inter-visibility matrix between each building, where the vertex points for buildings are points A-J, the points for the auxiliary line are points K-M, and the vector following the auxiliary line $ \mathbf{v_1}-\mathbf{v_3}  $.
	There are two main scenarios for computing the angular bouncing range in actual construction.
	The first scenario is when two faces are parallel to each other and the second scenario is when two faces are perpendicular to each other.
	These two scenarios are shown in Fig. \ref{fig:sysmodelANGover180}(b) and Fig.~\ref{fig:sysmodelANGover180}(c), respectively.	
	They serve as examples to illustrate how the algorithm functions and can also be applied to various other situations between buildings. 
	The interface relation in the inter-visibility matrix records the parallel or perpendicular scenarios and the xy-plane or yz-plane.
	
	Fig. \ref{fig:sysmodelANGover180}(b) displays the inter-visibility information of faces AB and EF with face AB serving as a reference, where face AB and face EF are set in a parallel scenario. 
	The angular bouncing range for face AB becoming visible to face EF is ($\omega_\mathrm{p1}$, $\omega_\mathrm{p2}$) for edge B and ($\mu_\mathrm{p1}$, $\mu_\mathrm{p2}$) for edge A.
	The visible angles $\omega_\mathrm{p1}$ and $\omega_\mathrm{p2}$ with edge B, and the visible angels $\mu_\mathrm{p1}$ and $\mu_\mathrm{p2}$ with edge A can be easily computed by the dot product of respective vectors $ \overrightarrow{\mathrm{AK}} $.
	Let compute $\omega_\mathrm{p1}$ as an example, 
	\begin{equation}\label{equ:dptproduct}
		\omega_\mathrm{p1}=\cos^{-1}{\left[ \frac{\overrightarrow{\mathrm{BA}} \cdot \overrightarrow{\mathrm{BF}}}{|\overrightarrow{\mathrm{BA}}||\overrightarrow{\mathrm{BF}}|}\right] }.
	\end{equation} 	
		
	Fig. \ref{fig:sysmodelANGover180}(c) shows the inter-visibility information of faces AC and EG. 
	Similar to the xy-plane situation, the angular bouncing range can be easily computed since the face AC and the face EI represent a perpendicular scenario. 
	For the parallel situation between the face AC and face EG, with the face AC as a reference, the angular bouncing range is ($\omega^{'}_\mathrm{e2}$, $\omega^{'}_\mathrm{e3}$) for edge A and ($\mu^{'}_\mathrm{e1}$, $\mu^{'}_\mathrm{e3}$) for edge C. 
	For the perpendicular situation between the face AC and face EI, with the face AC as a reference, the angular bouncing range is ($\omega^{'}_\mathrm{e1}$, $\omega^{'}_\mathrm{e2}$) for edge A and ($\mu^{'}_\mathrm{e1}$, $\mu^{'}_\mathrm{e2}$) for edge C. 
	However, in the real world, the part of face AC that is close to edge C cannot be visible to face EI because the face AC is blocked by face EG.
	The inter-visibility matrix with reflection information for Fig. \ref{fig:sysmodelANGover180} is shown in Table~\ref{table:inter-matrix}, where parallel and perpendicular are denoted by PAR and PER.
	
	Moreover, in Fig. \ref{fig:sysmodelANGover180}(c), the values of $\omega_\mathrm{e1}$ and $\mu_\mathrm{e1}$ are less than 180 degrees, and the value of $\omega_\mathrm{e2}$ and $\mu_\mathrm{e2}$ will exceed 180 degrees, where the reference line is the $\overrightarrow{\mathrm{IL}}$ and the angle is measured clockwise.
	However, the visibility angle should be less than 180 degrees because the horizontal face can only be visible to the two sides of one vertical face.
	In our case, only the side facing away from the objects can be set as the visible face. 
	When two faces are perpendicular to each other in the propagation scenarios, the ray cannot propagate from the backward side of the vertical face to the horizontal face through the object.
	Accordingly, the reference line in perpendicular scenarios is suitable for considering the horizontal face, and only the visibility angle less than 180 degrees will be recorded.
	Angles over 180 degrees will be recorded as 0 degrees, coinciding with the reference line.
	The 0 degree angular information can be used to compute the interaction point of the horizontal plane to check the angular bouncing range. 
	Let compute $ \omega_{\mathrm{e}_i} $ as an example, where $ i \in \left\lbrace 1, 2, ... \right\rbrace $ is the number of the angular bouncing range for different points at one edge,
	\begin{equation}\label{equ:angularrange}
			\omega_{\mathrm{e}_i} = \left\{ \begin{array}{ll} 
			\omega_{\mathrm{e}_i} , &\omega_{\mathrm{e}_i} < 180^{\circ},\\
			0, & \ 180^{\circ} \le \omega_{\mathrm{e}_i}.
		\end{array} \right.
	\end{equation}
	For some specific cases, the overlap situation between two buildings for single reflection case and how judgment the visibility between transmitter and reference face are mentioned in following Sections.
	
	\begin{table}[t]
		\centering
		\caption{Inter-visibility relationship of each face in Fig. \ref{fig:sysmodelANGover180}.}
		\begin{tabular}{|c|c|c|c|}
			\hline
			Order &Reference Face &Aim Face & Interface Relations \\
			\hline
			1  &AB &EF &PAR \\
			\hline
			1  &EF &AB &PAR \\
			\hline
			1 &FG &HJ  &PAR\\
			\hline
			1  &FG &HI &PER\\
			\hline
			1  &HI &FG &PER\\
			\hline
			1  &HJ &FG &PAR\\
			\hline
		\end{tabular}
		\label{table:inter-matrix}
	\end{table}
	\subsection{Multiple Reflections Procedure for Static Status}
\begin{figure*}[t]
		\centering
		\subfloat[xy-plane]{\includegraphics[width=0.33\textwidth]{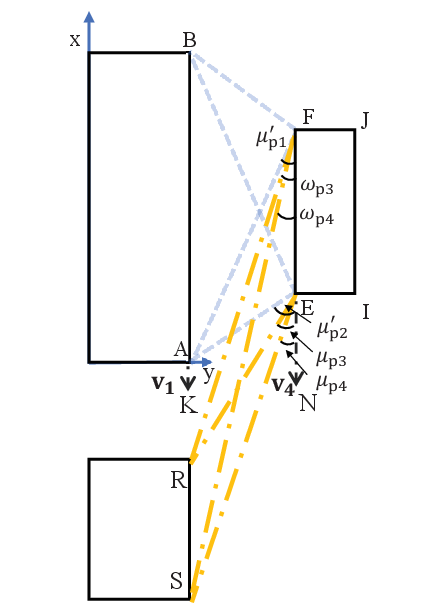}}
		\hfil
		\subfloat[xy-plane (overlap situation)]{\includegraphics[width=0.33\textwidth]{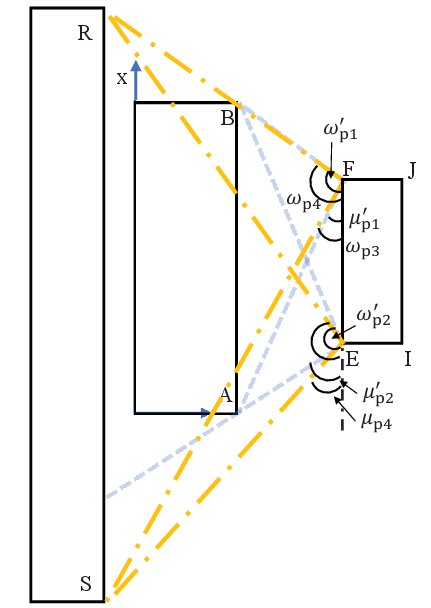}}
		\hfil
		\subfloat[yz-plane]{\includegraphics[width=0.33\textwidth]{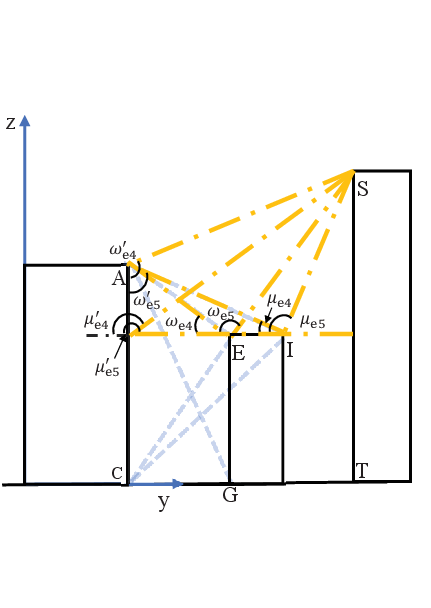}}
		\hfil
		\caption{Determination visibility angles of a visible face for multiple reflection in (a) xy-plane view, (b) xy-plane view with overlap situation, and (c) yz-plane. The blue dash lines express the original angular bouncing range, and the yellow long dash dor line express the second-order angular bouncing range.}
		\label{fig:sysmodelmulti}
	\end{figure*}

	In the case of multiple reflections, the reflection depends on the relative position of the nearby building.
	The interface relation and angular bouncing range between each building in the considered area will be used as criterion to compute the multiple reflections. 
	A third building is added in Fig.~\ref{fig:sysmodelANGover180} to explain how the static buildings for multiple reflections are regenerated in Fig. \ref{fig:sysmodelmulti}, wherein the angular bouncing range of the visible face in Fig. \ref{fig:sysmodelANGover180} has been ignored to improve clarity.
	The angular bouncing ranges in Fig. \ref{fig:sysmodelANGover180} and Fig. \ref{fig:sysmodelmulti} are the same.	
	For simplicity, second-order reflection cases for parallel and perpendicular situations are considered as examples, and multiple reflections could follow the proposed method.
	
%

	In Fig \ref{fig:sysmodelmulti}(a), when face AB is the original reference, face EF is the first-order reflection and face {\color{blue}RS} is the second-order reflection for the point on face AB, respectively.
	The angular bouncing range for the face EF becoming visible for the face {\color{blue}RS} can be computed using same method in Section II-B.
	The visible angles with edge F are $\omega_\mathrm{p3} $ and $\omega_\mathrm{p4} $, and the visible angles with edge E are $\mu_\mathrm{p3} $ and $ \mu_\mathrm{p4}$ for face {\color{blue}RS}.
	Then, the criterion for justifying whether second-order reflection occurs is defined by the relative angular information with the first-order reflection angle.
	Since face AB and EF are parallel, relative angular information can be computed by using the same simple mathematics principle, e.g., $\mu^{'}_\mathrm{p1}=180-\mu_\mathrm{p1} $.
	The example conditions in mathematical expression for the face {\color{blue}RS} (test face) becoming visible with second-order reflection as the face AB are shown below
	\begin{equation}\label{equ:secondref_con}
		\begin{aligned}
			\mu^{'}_\mathrm{p1} \geq \omega_\mathrm{p3} &\cap \mu^{'}_\mathrm{p1} \geq \omega_\mathrm{p4}  \\
			\mu^{'}_\mathrm{p2} \geq \mu_\mathrm{p3} &\cap \mu^{'}_\mathrm{p2} \geq \mu_\mathrm{p4}.
		\end{aligned}
	\end{equation}
	Without loss of the generality, the same method can be used for any location of the test face is relative to the original face, where the $\mu^{'}_\mathrm{p1}$ and $\mu^{'}_\mathrm{p2}$ can be replaced by $\omega^{'}_\mathrm{p1}$ and $\omega^{'}_\mathrm{p2}$.
	In this situation, the criterion condition should be changed with the relative position with simple geometric mathematics.
	When both conditions are true, the whole area of test face can be defined as the second-order reflection of the reference face.  
    If the condition does not all true, it means the test face has an overlap area with the reference face, which is shown in Fig.~\ref{fig:sysmodelmulti}(b).
    The interface relation in matrix will label the face relative position information and separated it into non-overlap and overlap parts.
    The angular bouncing range of the non-overlap part will be computed and recorded with the criterion angular and angular bouncing range. 
	
	In Fig. \ref{fig:sysmodelmulti}(c), there are two cases for the second-order reflection. 
	Case 1 is the vertical to horizontal to vertical face and case 2 is horizontal to vertical to vertical. 
	Other cases can be simplified to these two cases because the order of the visibility face is reversible.
	For case 1, the second-order reflection can be divided into two first-order reflection processes, because two vertical faces will not overlap with the same horizontal face.
	Accordingly, the angular bouncing range of vertical to horizontal face and horizontal to vertical face can be recorded directly in the 3D inter-visibility matrix for first-order reflection.
	The interface relation for these faces will record the order of face connecting. 
	For case 2, the horizontal to vertical part can be computed with the same method as case~1.
	Moreover, the vertical to vertical part can be computed as a parallel situation, where these two vertical faces must be separated and non-overlap.
	The interface relation will record the order of face connecting and the face type for vertical and horizontal. 
	The main idea is to consider the limitation between each face in building as the criterion and the original angular bouncing range together as a whole condition to build the inter-visibility matrix.
	These methods can be iterated in a similar way for more than second-order reflection cases.
	
	\section{3D Inter-visibility Table for Mobile Status}
	The 3D inter-visibility matrix describes the visibility relationship among static buildings in the scenario, which is a mid-process of the 3D inter-visibility table for the moving transmitter and receiver.
	Then, the moving trajectories of the transmitter and receiver will be added to the static scenario, where the inter-visibility matrix will extend to the inter-visibility table.
	How the inter-visibility table generation, which is used to find the ray creation or suppression points, will be introduced in this section.
	 
	\subsection{Mobile Source in Inter-visibility Matrix }
\begin{figure}[t]
		\centering
	\subfloat[xy-plane]{\includegraphics[width=0.33\textwidth]{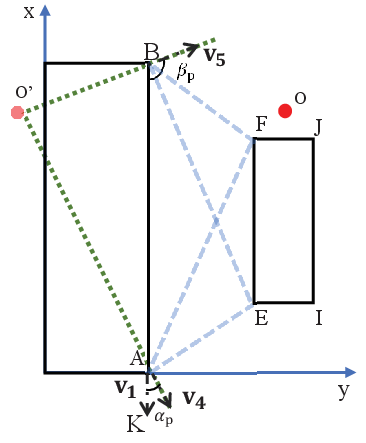}}
	\hfil
	\subfloat[yz-plane]{\includegraphics[width=0.33\textwidth]{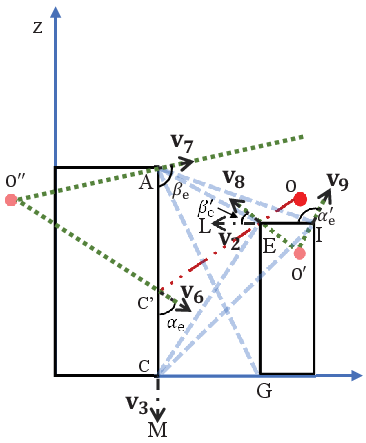}}
	\hfil
	\caption{Determination reflection order of a visible face in (a) xy-plane view and (b) yz-plane view. The blue dash lines express the angular bouncing range, the green dot lines express the visible angular range for reflection, and the red long dash dot dot line express the visible angular range of transmitter.}
	\label{fig:sysmodelref}
\end{figure}

	When a mobile transmitter is added as a source point into the static scenario, reflected rays for the source point are produced within a specific angular range, referred to as the illuminated region of the image point.
	A modified version of Fig. \ref{fig:sysmodelANGover180} is presented to explain how the image point for reflection is regenerated in Fig. \ref{fig:sysmodelref}.
	The angular bouncing range of the visible face has been ignored to improve the clarity, but the angular bouncing ranges in Fig. \ref{fig:sysmodelANGover180} and Fig. \ref{fig:sysmodelref} are the same.
	
	In Fig. \ref{fig:sysmodelref}(a), the transmitter point and image transmitter point with face AB are denoted as O and O'.
	The visible angular range $ \alpha_\mathrm{p}$ and $\beta_\mathrm{p} $ for the first-order reflection with face AB can be computed by vector $\mathbf{v}_1$, $\mathbf{v}_4 $, and $\mathbf{v}_5 $ by dot product.
	The conditions in mathematical expression for face EF (aim face) to make a valid higher-order visible node of face AB (reference face) in Fig. \ref{fig:sysmodelref}(b) are expressed as
	\begin{equation}\label{equ:conditionp}
		\begin{aligned}
			\alpha_\mathrm{p} < \mu_\mathrm{p1} &\cap \beta_\mathrm{p} >\omega_\mathrm{p1}  \\
			\alpha_\mathrm{p} < \mu_\mathrm{p2} &\cap \beta_\mathrm{p} >\omega_\mathrm{p2}.
		\end{aligned}
	\end{equation}
	Furthermore, the reference face means the face with the reflection point, and the aim face means the illuminated face by reflection ray.  
	\begin{figure*}[t]
	\centering
		\subfloat[$ \alpha_\mathrm{p} < \mu_\mathrm{p1} \cap \beta_\mathrm{p} <\omega_\mathrm{p1} $]{\includegraphics[width=0.35\textwidth]{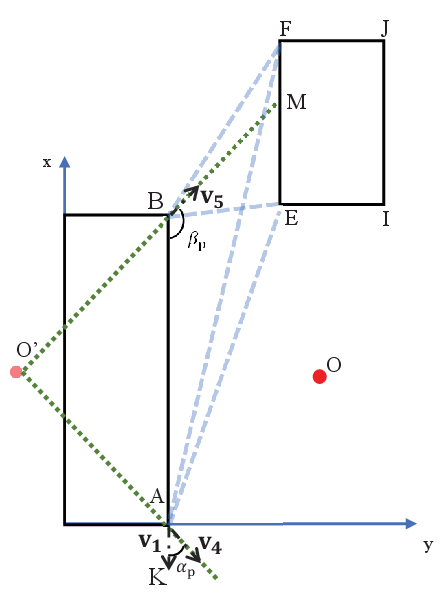}}
	\hfil
	\subfloat[$ \alpha_\mathrm{p} > \mu_\mathrm{p2} \cap \beta_\mathrm{p} >\omega_\mathrm{p2} $]{\includegraphics[width=0.35\textwidth]{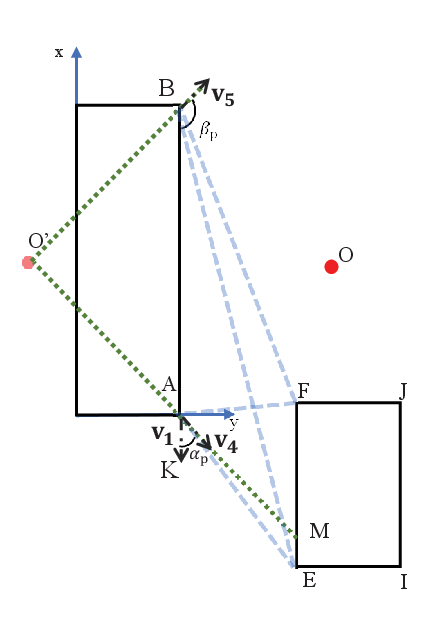}}
	\hfil

	\caption{Determination reflection order between a face and a part visible face in xy-plane view for (a) the aim building over the reference building and (b) the aim building under the reference building . The blue dash lines express the angular bouncing range and the green dotted lines express the visible angular range for reflection.}
	\label{fig:sysmodelpart}
\end{figure*}

	For more general cases, when both conditions are true, it indicates that the entire area of aim face is visible to reference face.
	In this case, the reflection angular range is defined with relative angular bouncing range.
	In contrast, if all conditions are false, the aim face is out of the reflection range with reference face.
	In addition, if only one condition is true, it means part of the aim face area is invisible to the reference face, which is shown in Fig. \ref{fig:sysmodelpart}.
	Consequently, the reflection angular range is determined by the angular bouncing range and the visible angular range, which can be computed using simple geometric relations.
	Since there are at least two points for visible angle line and the aim face are known, the general line function can be given as slope-intercept form $ y = mx + n$, where $ m $ is the slope of the line and $ n $ is the y-value of the y-intercept.
	The visible angle line is used to check the visible angular range between the aim face and reference face, when part of the aim face area is invisible to the reference face.
	The intersection point M for visible angular range and aim face can be computed by
	\begin{equation}\label{equ:point}
		\mathrm{M}=\left( \frac{n_v-n_a}{m_a-m_v}, \frac{m_an_v-m_vn_a}{m_a-m_v} \right), 
	\end{equation}
	where $m_v $ and $n_v$ denote the slope and intercept of visible angle line, and $m_a $ and $n_a$ denote the slope and intercept of aim face, respectively.
	
	In Fig. \ref{fig:sysmodelref}(b), the transmitter point with height has two image points with face EI (horizontal plane) and face AC (vertical plane), denoted O' and O'', respectively.
	In general, when the transmitter point can be illustrated to the whole reference face, the computation procedure is similar to the previous described cases in the xy-plane.
	However, the reference face only shows part of the area with transmitter point in some cases because of the blockage by the nearby building.  
	Accordingly, the visible angle line needs to be used to compute the new angular bouncing range, where the line OC’ is the visible angel line in Fig. \ref{fig:sysmodelref}(b).
	The new visible angular range with point C' can be computed by finding the intersection of reference face and transmitter point following (5). 
	For mathematical expression, the conditions for face AC to make a valid higher-order visible node of face EG are expressed as
	\begin{equation}\label{equ:conditione'}
		\begin{aligned}
			\alpha^{'}_\mathrm{e} < \mu^{'}_\mathrm{e1} &\cap \beta^{'}_\mathrm{e} >\omega^{'}_\mathrm{e1}  \\
			\alpha^{'}_\mathrm{e} < \mu^{'}_\mathrm{e3} &\cap \beta^{'}_\mathrm{e} >\omega^{'}_\mathrm{e3}.
		\end{aligned}
	\end{equation} 
	Similarly, the conditions for face EI to make a valid higher-order visible node of AC are expressed as 
	\begin{equation}\label{equ:conditione}
			\alpha_\mathrm{e} < \mu_\mathrm{e1} \cap \beta_\mathrm{e} >\omega_\mathrm{e1}.
	\end{equation}
	The reason for only one condition is that $ \alpha_\mathrm{e2}$ and $\beta_\mathrm{e2} $ are set as 0 degrees, where only part of the aim face can be visible to the reference face.
	
	\subsection{3D Inter-visibility Table for Specific Points}
	Before generating a 3D inter-visibility table for mobile status, the inter-visibility table for specific points should be considered because this table is the basis part of the mobile status. 
	Based on the visibility relationship, the parent node transmitter searches the potential reference faces for reflection. 
	This ensures that only the appropriate faces are considered for further processing, reducing unnecessary computations and improving overall efficiency.
	The inter-visibility matrix offers information on the aim face for reflection or other propagation mechanisms based on the setting reference face. 
	After computing the angular bouncing range and comparing it with the inter-visibility matrix, the 3D inter-visibility table for specific points can be generated.
	
	For Fig. \ref{fig:sysmodelref}, the transmitter O is visible to face AC, face AC', and face EI, so these faces will be set as the reference faces.
	Based on the inter-visibility matrix, aim faces related to these reference faces can be found. 
	Accordingly, the inter-visibility table for Fig. \ref{fig:sysmodelref} can be computed, which is shown in Table \ref{table:1}. 
	It should be noted that the inter-visibility table is combined by two inter-visibility tables of xy-plane and yz-plane together.  
	\begin{table}[t]
		\centering
		\caption{Inter-visibility table of each face in Fig. \ref{fig:sysmodelref}.}
		\begin{tabular}{|c|c|c|c|}
			\hline
			Order &Parent Node &Reference Face &Aim Face \\
			\hline
			1 &transmitter &AB &EF  \\
			\hline
			1 &transmitter &AC' &EG  \\
			\hline
			1 &transmitter &AC' &EI \\
			\hline
			1 &transmitter &AC' &CG  \\
			\hline
			1 &transmitter &EI &AC' \\
			\hline
		\end{tabular}
		\label{table:1}
	\end{table}
	
	 For the specific receivers, the first-order diffraction can be supported in the proposed method. 
	The process of finding the closest edge for the receiver is similar to finding the reflection face, where the vertex points can be used to define the face and the edge. 
	We incorporate diffraction by identifying the closest edge for the receiver in the inter-visibility matrix relevant to computing first-order diffraction. 
	Then, the UTD is applied to calculate the power of diffraction propagation, incident angles, and diffracted angles. 
	Since the angle of diffraction depends on the edge of the visible face for receiver and the visible face for reflection and diffraction is same, the angle of diffraction does not impact the 3D inter-visibility table. 

	\subsection{3D Inter-visibility Table for Mobile Status}
	For generating the 3D inter-visibility table for mobile status, inter-visibility tables of specific points along the route should be computed.
	These data points along the route can be used to find the visibility change node as the visibility range criterion. 
	The yz-plane information should be justified firstly for the inter-visibility table because the vertical position information projects to horizontal plane easily.
	The velocity of transmitter and receiver is used to generate their moving trajectories, which is used to build the inter-visibility table and determine the visibility range.

\begin{figure}[t]
		\centering
		\subfloat[xy-plane]{\includegraphics[width=0.45\textwidth]{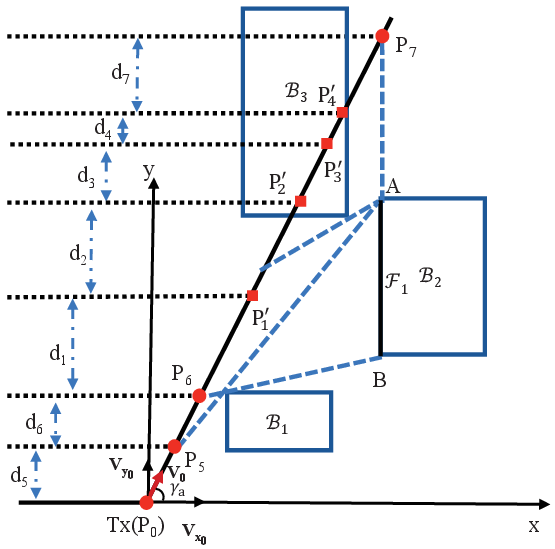}}
		\hfil
		\subfloat[yz-plane]{\includegraphics[width=0.45\textwidth]{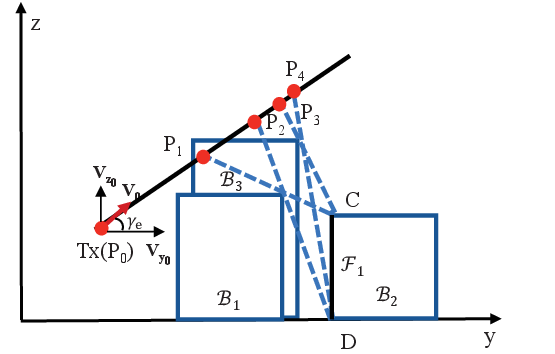}}
		\hfil
		\caption{Visible regions of face $ \mathrm{F}_1 $ for a mobile transmitter in (a) xy-plane view and (b) yz-plane view.}
		\label{fig:VisReg}
	\end{figure}

	The example of the transmitter moving with visibility to face $ \mathcal{F}_1 $ is shown in Fig. \ref{fig:VisReg}, where the height changing of the transmitter in the yz-plane can be projected to the transmitter route in xy-plane.
	$ \mathrm{V_{x_0}} $, $ \mathrm{V_{y_0}} $, and $ \mathrm{V_{z_0}} $ are the reference vectors parallel with x-axis, y-axis, or z-axis to compute the direction of the moving transmitter in Fig. \ref{fig:VisReg}.
	The projected distance of the xy-plane can be computed by the height change and direction with speed information.
	The visible nodes for face $ \mathcal{F}_1 $ are points A, B, C, and D. 
	The yz-plane has 4 visible nodes related to the buildings $ \mathcal{B}_1 $ and $ \mathcal{B}_3 $ considering face CD as reference, which can be projected to the xy-plane.
	The yz-plane visibility table can be generated easily following the previous method, where the  visibility range points 1-4 for the yz-plane are shown in Fig.~\ref{fig:VisReg}(b) and the visibility table is shown in Table \ref{table:2}. 
	The final visibility table is generated and illustrated in Table \ref{table:3} and the all visibility range segments can be found in Fig. \ref{fig:VisReg}(a), where the projection visibility range points in yz-plane are denoted by $ \mathrm{P}^{'} $.

	The inter-visibility table of multiple reflection for the mobile transmitter can be generated based on the multiple reflection inter-visibility matrix.
	The source point and the reflection face can produce the first-order reflection angular bouncing range as reference angular information, which will transfer to the inter-visibility matrix.
	The multiple-order information about the static buildings can support the generation of higher-order reflection information based on the reference angular information.
	The chosen specific points depend on the reflection order, which contains the main transmission power of electromagnetic wave propagation. 
	Since the main transmission power is contained in free space, first-order reflection, and second-order reflection propagation \cite{Cuinas2020APM}, the specific points along the route are computed for these propagation mechanisms. 
	The coherence time accuracy is related to the specific points generated following the different orders of reflection.
		
	\begin{table}[t]
		\centering
		\caption{Visibility table of each face in the yz-plane.}
		\begin{tabular}{|c|c|c|c|c|}
			\hline
			 Order & Visible Node & Visibility Range& Parent Node & Blockage \\
			\hline
			1 & $ \mathcal{F}_1 $ &$\mathrm{P}_1-\mathrm{P}_2$  & transmitter & $\mathcal{B}_1$\\
			\hline
			1 &$ \mathcal{F}_1 $ &$\mathrm{P}_3-\mathrm{P}_4$ & transmitter & $\mathcal{B}_3$ \\
			\hline
			1 &C &$\mathrm{P}_1-\mathrm{P}_4$ & transmitter & $\mathcal{B}_1$ \\
			\hline
			1 &C &$\mathrm{P}_3-\mathrm{P}_4$ & transmitter & $\mathcal{B}_3$ \\
			\hline
			1 &D &$\mathrm{P}_2-\mathrm{P}_4$ & transmitter & $\mathcal{B}_1$ \\
			\hline
			1 &D &$\mathrm{P}_4-\mathrm{P}_4$ & transmitter & $\mathcal{B}_3$ \\
			\hline
		\end{tabular}
		\label{table:2}
	\end{table}

	\begin{table}[t]
		\centering
		\caption{Visibility table for the example scenario.}
		\begin{tabular}{|c|c|c|c|c|}
			\hline
			Order  & Visible Node & Visibility Range& Parent Node&Blockage \\
			\hline
			1 & $ \mathcal{F}_1 $ &$\mathrm{P}_5-\mathrm{P}_7$  & transmitter& $\mathcal{B}_1$\\
			\hline
			1 & $ \mathcal{F}_1 $ &$\mathrm{P}^{'}_3-\mathrm{P}_7$  & transmitter& $\mathcal{B}_3$\\
			\hline
			1 &A &$\mathrm{P}_5-\mathrm{P}_7$ & transmitter& $\mathcal{B}_1$\\
			\hline
			1 &A &$\mathrm{P}^{'}_3-\mathrm{P}_7$ & transmitter& $\mathcal{B}_3$\\
			\hline
			1 &B &$\mathrm{P}_6-\mathrm{P}_7$ & transmitter& $\mathcal{B}_1$\\
			\hline
			1 &B &$\mathrm{P}^{'}_4-\mathrm{P}_7$ & transmitter& $\mathcal{B}_3$\\
			\hline
		\end{tabular}
		\label{table:3}
	\end{table}

\begin{figure}
		\centering
		\includegraphics[width=0.88\linewidth]{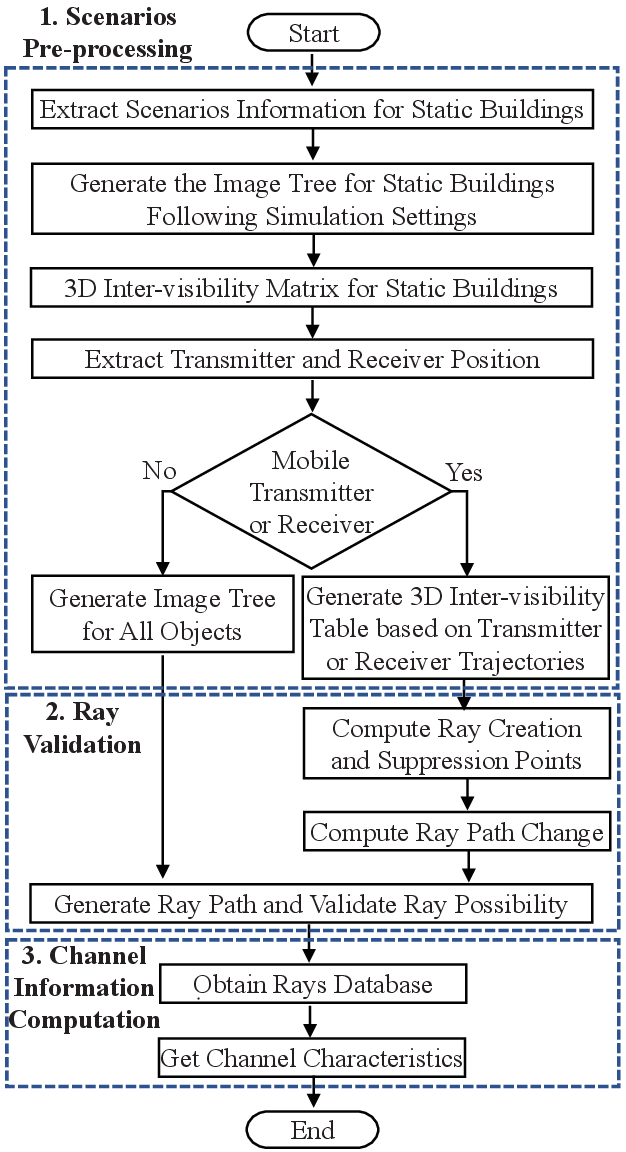}
		\caption{Flow chart of the ray-tracing method with 3D inter-visibility pre-processing. }
		\label{fig:Flowinv}
	\end{figure}

	The flow chart about the proposed pre-processing steps for the dynamic ray-tracing method is illustrated in Fig. \ref{fig:Flowinv}. 
	Firstly, the scenario information contains detailed geometric and material information for the ray-tracing simulations, which contains the position of all vertex points for the building, face ID, edge ID, material properties, etc.. 
	The sequence and application of the inter-visibility matrix and	inter-visibility table are explained. 
	The 3D inter-visibility table is important, enabling dynamic ray-tracing to compute ray creation and suppression, and avoiding unnecessary traversal of all scenario faces.	
	Additionally, the dynamic ray-tracing process is applied during the ray validation step to accurately compute changes in the ray path, ensuring precise and efficient simulations in complex environments.

	\subsection{Coherence Time Computation}
	Based on the 3D visibility table, the coherence time of the scenario could be computed.
	In the visibility table, the visibility range for each visible node and parent node has been computed to show each ray creation and suppression.
	The coherence time is the time duration over which the signal envelope remains constant, so it could be considered as no creation and suppression of the ray in the dynamic ray-tracing.
	Accordingly, the different visible range can be used to compute the coherence time of visibility range $ l $ as follows
	\begin{equation}\label{eq:coherencetime_s}
		T_{\mathrm{c},l}=\frac{d_{l}}{v_{l}},
	\end{equation} 
	where $d_{l}$ is the reality route distance of visibility range $l$, $ l \in \{1, 2, ..., L\} $, $L$ is maximum number of visibility range, and $v_{l} $ is the velocity of mobile objective in the visibility range~$ l $. Moreover, $ d_{l} $ in Fig. \ref{fig:VisReg} is computed by double projection from the 3D distance to a one-directional distance based on the angle between the reference velocity vector and velocity $ \gamma_\mathrm{e} $ and $ \gamma_\mathrm{a} $.
	Accordingly, the ray computation in this stable period can only compute the ray change information and not trace the ray again because there is no new ray creation or suppression \cite{Bilibashi2023TAP}.
	The most power of electromagnetic propagation is contained in direct, first-order to third-order reflection rays, so the ray creation and suppression for visibility range to compute coherence time is less third order reflection in our 	simulation \cite{He2019TUT}.
	The $T_{\mathrm{c},l}$ can be used to support the ray-evolution algorithm in dynamic ray-tracing method \cite{Bilibashi2023TAP}.
	
	There are two typical cases for computing the coherence time for moving objectives.
	For one objective moving case, the visibility range and velocity are computed based on the moving objective to use compute the coherence time, where \eqref{eq:coherencetime_s} can be used directly in this case.
	Additionally, the coherence time for transmitter and receiver side will be computed by the visibility range and velocity of transmitter and receiver in two objectives moving case, respectively.
	Moreover, the smallest coherence time between transmitter side and receiver side is considered as coherence time for this visibility range. 
	For statistical meaning, the average coherence time of the whole transmission period can be computed as
	\begin{equation}\label{eq:coherencetime}
		T_{\mathrm{c}}=\frac{\sum_{l=1}^{L}{T_{\mathrm{c},l}}}{L}.
	\end{equation}
	
	\section{Results and Analysis}
	The proposed pre-processing method is validated by two case studies, because it could be used in the static scenario and dynamic scenario and two modes should be validated, respectively. 
	In case 1, a realistic suburban scenario is considered, and simulation results are compared with measurement, where the 3D inter-visibility matrix and the image tree generating process of the proposed pre-processing method for static building are validated.
	In case 2, the simulation results are obtained in a simple 3D mobile transmitter environment by traditional and proposed pre-processing method. 
	The accuracy and efficiency of the proposed pre-processing method are evaluated by comparing these two methods.
	
	\subsection{Case Study 1: Comparison with Measurement}
\begin{figure}
		\centering
		\includegraphics[width=0.8\linewidth]{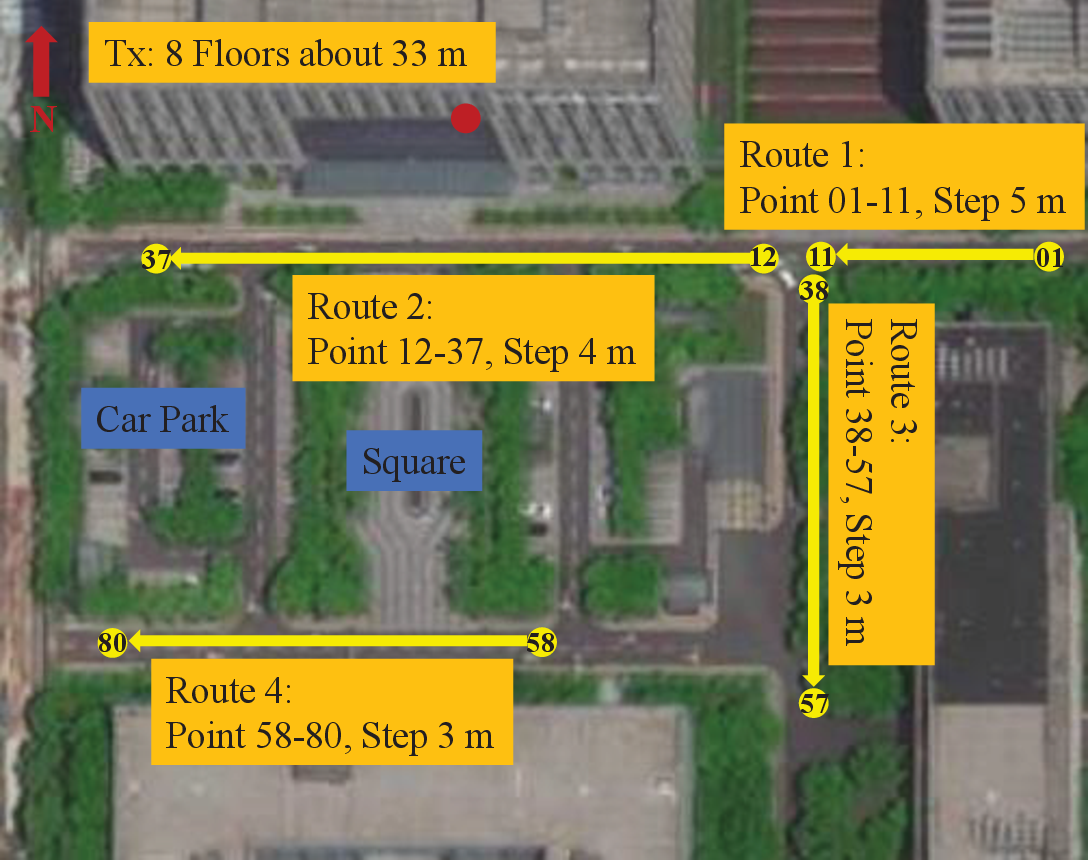}
		\caption{The outdoor channel measurement scenarios.}
		\label{fig:MeaCam}
	\end{figure}

	We perform the validation of proposed ray-tracing simulation results for a suburban scenario by comparing with channel measurement data, i.e., path loss, root-mean-square (RMS) delay spread. 
	 The maximum third-order reflection and one diffraction are used in the ray-tracing algorithm.
	The measurement area is about 200 m $ \times $ 100 m, which is shown in  Fig. \ref{fig:MeaCam}.
	The area contains a square, a car park area, and intersections, which are the main components of the urban scenario. 
	Furthermore, foliage, i.e., trees, in the measurement area leads to the NLoS propagation situation, which has been considered in our simulation. 
	
	The measurements were conducted at 5.5 GHz with a bandwidth of 320 MHz. 
	The transmitter side used an omnidirectional antenna with gain of 3 dBi and the receiver side was equipped with a 64 elements antenna array. 
	With the receiver antenna mounted on the trolley with a height of about 1.5 m \cite{Zheng2023TVT}. 
	The transmitter was located on the 8th-floor of a building with a height of about 33 m, remaining static throughout the measurements. 
	The receiver was moved point by point along the measurement routes, as shown in Fig. \ref{fig:MeaCam}. 
	Four measurement routes, containing a total of 80 receiver points, were conducted in a typical suburban scenario to validate the proposed pre-processing and ray-tracing algorithm.
	 
\begin{figure}
	 	\centering
	 	\includegraphics[width=0.95\linewidth]{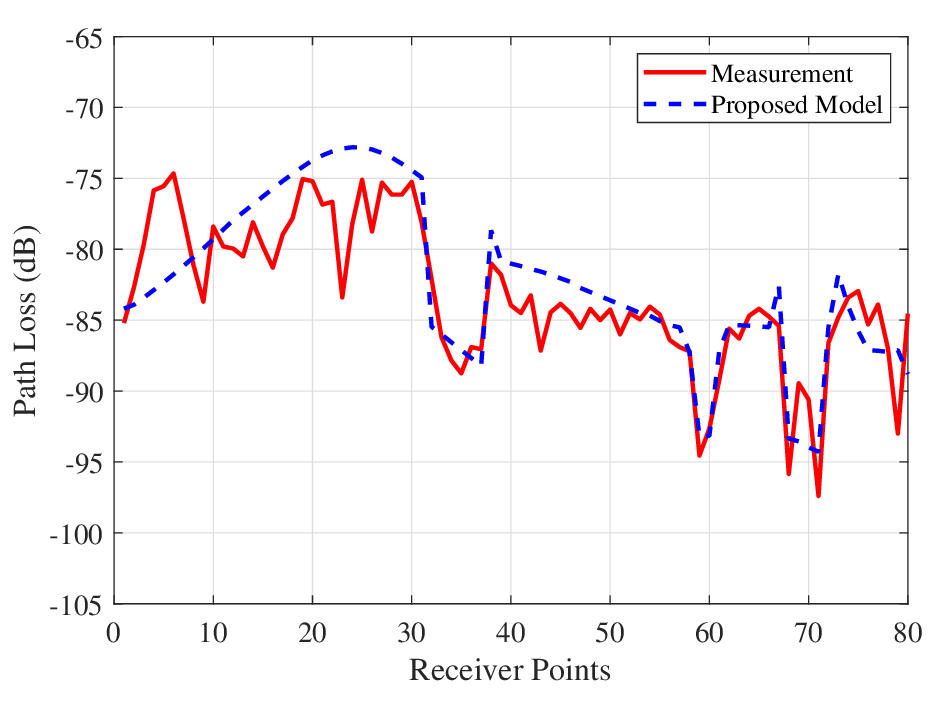}
	 	\caption{Comparison of path losses between measurement and simulation.}
	 	\label{fig:Pl}
	 \end{figure}

	 The measured and ray-tracing simulated path loss with LoS and NLoS are shown in Fig. \ref{fig:Pl}. 
	 The transmission power is set at 0 dBm in ray-tracing simulation because the measurement system was calibrated before measurement and the calibrated data will be used to normalized the transmission power.
	 In general, the value of path loss is a good agreement between measurement and simulation results.
	 The mean absolute error and RMS error of path loss between measurement and ray-tracing simulation results are 2.44 dB and 3.10 dB, respectively.
	  Many scatterers, e.g., street lamps and cars, are not modeled in our simulation environment, so these missing scatterers cause the difference in path loss.
	 The receiver points 59, 60, 69-72, and 75-79 are NLoS points, which are blocked by the foliage in the square. 
	 The value of ray-tracing simulated path loss at these points is attenuated heavily, which is also in good agreement with the measurement.
	 
\begin{figure}
 		\centering
 		\includegraphics[width=0.65\linewidth]{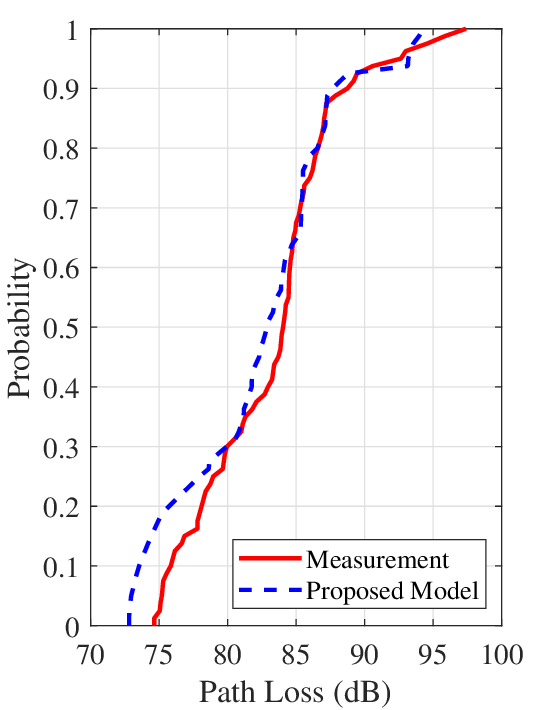}
 		\caption{CDFs of path losses between measurement and simulation.}
 		\label{fig:EFFPl}
 	\end{figure}

\begin{figure}
 	\centering
 	\includegraphics[width=0.65\linewidth]{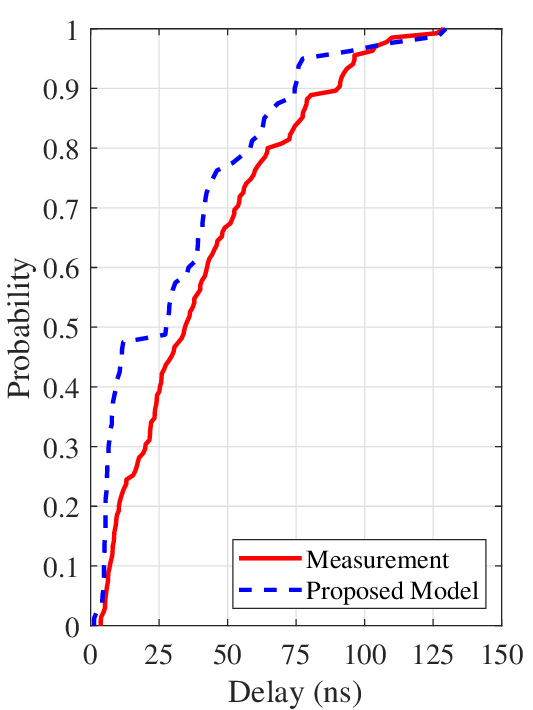}
 	\caption{CDFs of RMS delay spreads between measurement and simulation.}
 	\label{fig:EFFDS}
 \end{figure}

	 The cumulative distribution function (CDF) of path loss and RMS delay spread is shown in Fig. \ref{fig:EFFPl} and Fig \ref{fig:EFFDS}.
	 The proposed ray-tracing simulation result fits in good agreement with the measurement results.
	 The multipath effect of the wireless channel can be expressed by the RMS delay spread. 
	 In our setting environment, many foliage, street lamps, and cars exist near the road and in the park, where these scatterers lead to the enriching multipaths in the environment.  
	 Accordingly, the larger RMS delay spread is observed in the measurement results compared to the proposed ray-tracing simulation results. 
	 The trend between measurement and simulation with CDF of RMS delay spread is similar, where the main scatterers, i.e., buildings, impacts on the ray propagation can be captured. 
	 These comparisons show the accuracy of the proposed method from the channel characteristics perspective.
 
	\subsection{Case Study 2: Ideal Urban Scenario}
\begin{figure}[t]
		\centering
		\subfloat[xy-plane]{\includegraphics[width=0.43\textwidth]{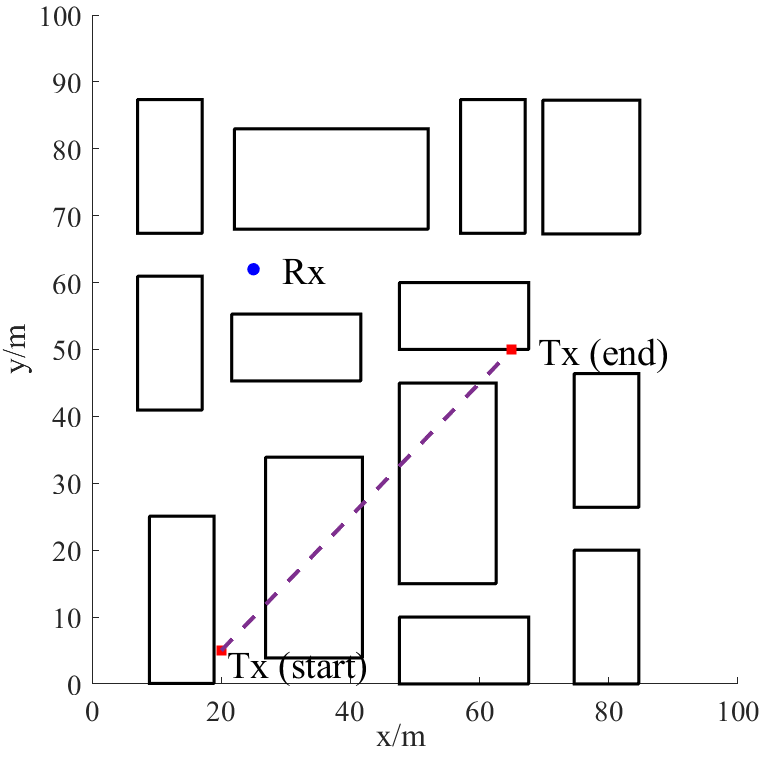}}
\hfil
\subfloat[yz-plane]{\includegraphics[width=0.43\textwidth]{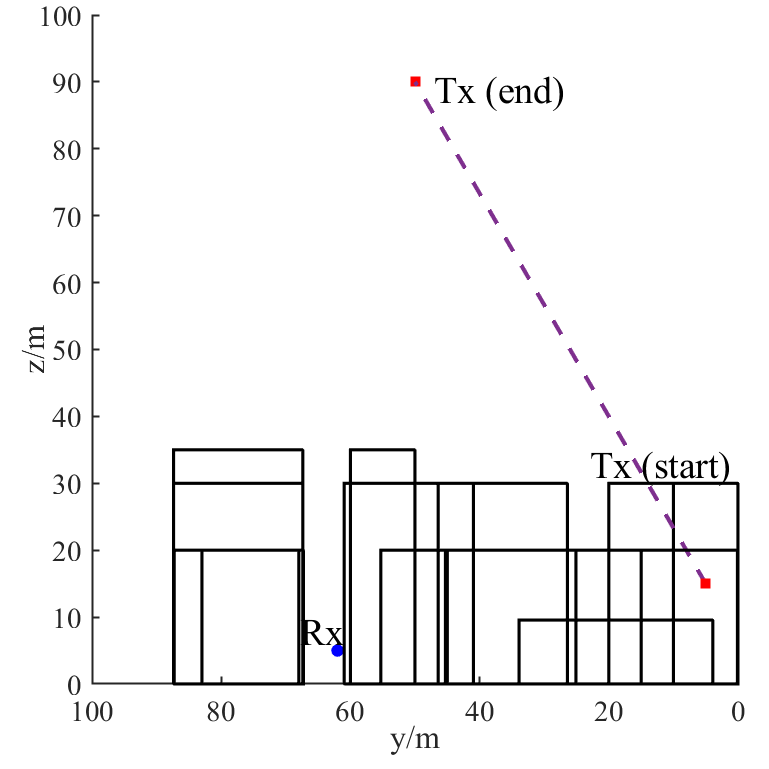}}
\hfil
		\caption{Simulation scenarios (a) xy-plane view and (b) yz-plane view for a mobile transmitter in typical urban scenarios.}
		\label{fig:Secnario}
	\end{figure}

	We perform a ray accuracy validation of the proposed pre-processing method in an ideal urban scenario by comparing it with the brute-force ray search method.
	The setting scenario is a Manhattan grid-like urban environment in 100 m$ \times $ 100~m square area.
	The scenario containing 13 buildings and the 3D mobile transmitter test trajectory is shown in Fig. \ref{fig:Secnario}. 
	Moreover, the total triangulation faces of case 2 are 158 after the Delaunay triangulation with 78 building faces and a ground face.
	This simulation scenario has considered some urban measurement scenario characteristics to validate the pre-processing method in an urban environment \cite{Hussain2019TAP, Boban2014TVT}. 
	The test trajectory of 120 transmitter points between (20, 5, 15) and (65, 50, 90) is considered, and the receiver point is located as (25, 62, 5), where the unit is a meter.  
	
\begin{figure}[t]
	\centering
		\subfloat[xy-plane]{\includegraphics[width=0.43\textwidth]{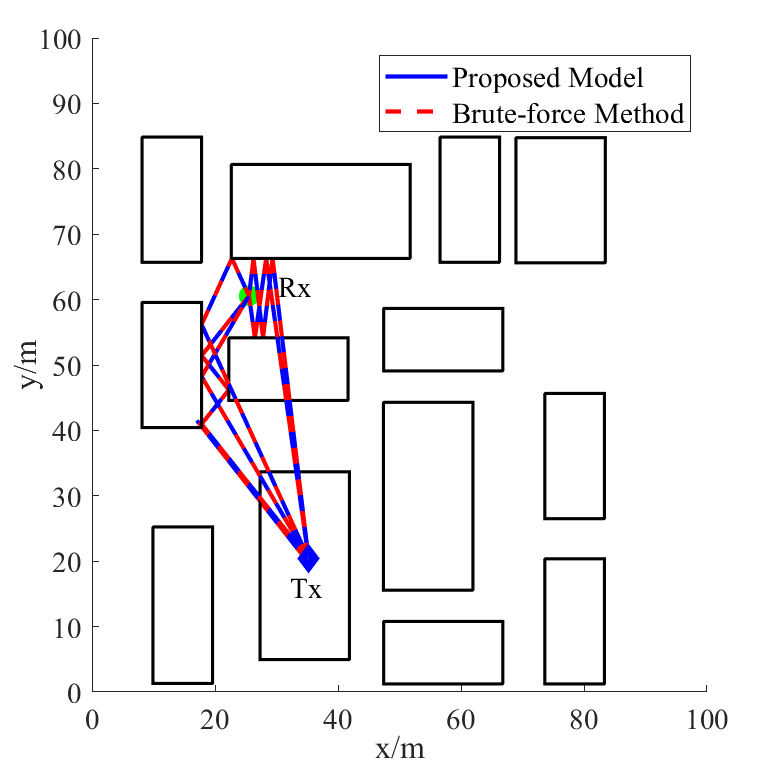}}
\hfil
\subfloat[yz-plane]{\includegraphics[width=0.43\textwidth]{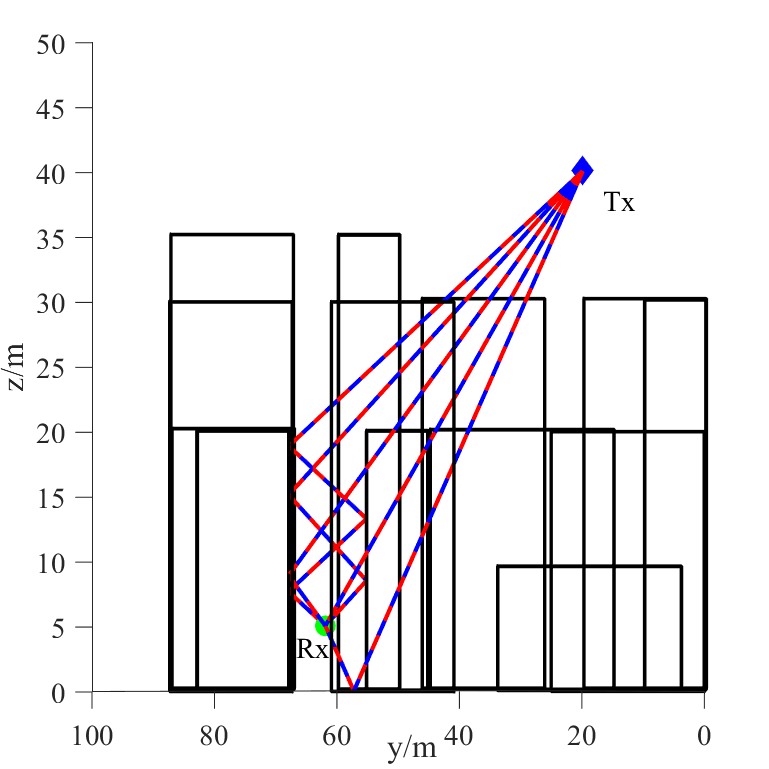}}
\hfil

	\caption{Comparison the accuracy of proposed model for searching rays against brute-force approach in (a) xy-plane view and (b) yz-plane view.}
	\label{fig:CompRT}
	\end{figure}

	We evaluate the accuracy of our proposed ray-tracing algorithm from the perspective of ray search. 
	We compute all the valid rays up to the third-order reflection (maximum one diffraction) for the transmitter and receiver pair. 
	Firstly, a brute-force approach is used to compute all the valid rays (up to the third-order reflection) for the given transmitter and receiver location.  
	The image tree is generated by the brute-force approach to compute valid rays at the receiver. 
	Similarly, we use the inter-visibility table to compute the valid rays for the transmitter trajectory to extract the image tree for the given transmitter locations.
	The valid rays arriving at the receiver from the transmitter are also computed using the proposed method. 
	We noted that the same valid rays are computed by these two different methods, as shown in Fig. \ref{fig:CompRT}. 
	The comparison of valid rays at the receiver for both methods indicates that rays computed using the proposed algorithm are the same as the ones obtained using the brute-force approach. 
	This comparison advocates the accuracy of the proposed model in ray search. 
	It should be noted that Fig. \ref{fig:CompRT} shows a snapshot of the channel for only a single transmitter-receiver pair, where the transmitter is (35, 20, 40) and the receiver is (25, 62, 5).
	This comparison shows the accuracy of the proposed method from the ray search perspective.

	\subsection{Processing Time Comparison}
	We provide a comparison of computation time for the same scenario for the proposed pre-processing ray-tracing method versus the traditional ray-tracing method verified in \cite{Lv2022ICCC}.
	In all simulations, a standard desktop with an Intel Core i7-10700 2.9 GHz CPU and 32 GB RAM was used.
	The time consumption of one ray-tracing run for one point with different reflection orders and the total time consumption of the whole trajectory with different reflection orders are shown in Table \ref{table:4}.
	From Table \ref{table:4}, we can observe computation time reduction using the proposed algorithm is up to 95\% for a fixed transmitter and receiver location compared to \cite{Lv2022ICCC}.
	
	\begin{table}[t]
	\centering
		\caption{Time consumption for different reflection-orders in test environments. }
		\label{table:4}
		\begin{tabular}{|c|c|c|c|c|}
			\hline
			Time (s)  & Order=1 & Order=2& Order=3&Order=4 \\
			\hline
			Proposed Method &1.718&1.835&1.897&15.234\\
			\hline
			Brute-force Method& 0.747&1.132&45.494&45442.607\\
			\hline
			\textbf{Reduction}&\textbf{-129\%}&\textbf{-28\%}&\textbf{95.8\%}&\textbf{99.96\%}\\
			\hline
		\end{tabular}
	\end{table}
	
	The proposed method generates the face-to-face inter-visibility table for up to fourth-order reflections for	the given test environment. 
	The processing time of the proposed and Brute-force method \cite{Lv2022ICCC} is shown in Table \ref{table:4}.
	The proposed model reduces the processing time for wall reflections by up to 95\% when the reflection order is over third-order.
	For first-order and second-order reflection, the processing time is slightly longer than the processing time in \cite{Lv2022ICCC} because the generation of the inter-visibility table involves extra processing time. 
	The generation of the inter-visibility table involves computing all faces and edges in the simulation scenario, whereas the traditional ray-tracing method only finds the visible faces according to the highest reflection order. 
	Hence, the proposed method spends more time in the pre-processing part at low reflection orders, but the cost time is similar. The difference between the proposed method and the traditional ray-tracing is within 1 s.
%
	\section{Conclusions}
	We have proposed a novel dynamic ray-tracing method that employs pre-processing to compute the moving objects in urban scenarios.
	Pre-processing has generated a 3D inter-visibility matrix for each face and edge of all the buildings in the environments.
	The 3D inter-visibility table has been developed based on the inter-visibility matrix to improve image tree efficiency.
	Simulation results have shown that the proposed ray-tracing method is considerably time-saving compared with the traditional method.
	The channel characteristics computed by our method closely align with channel measurements.  
	The proposed pre-processing method will enable the ray-tracing method in 6G high mobility applications, such as vehicular communications and UAV communications. 

\begin{IEEEbiography}[{\includegraphics[width=1in,height=1.25in,clip,keepaspectratio] {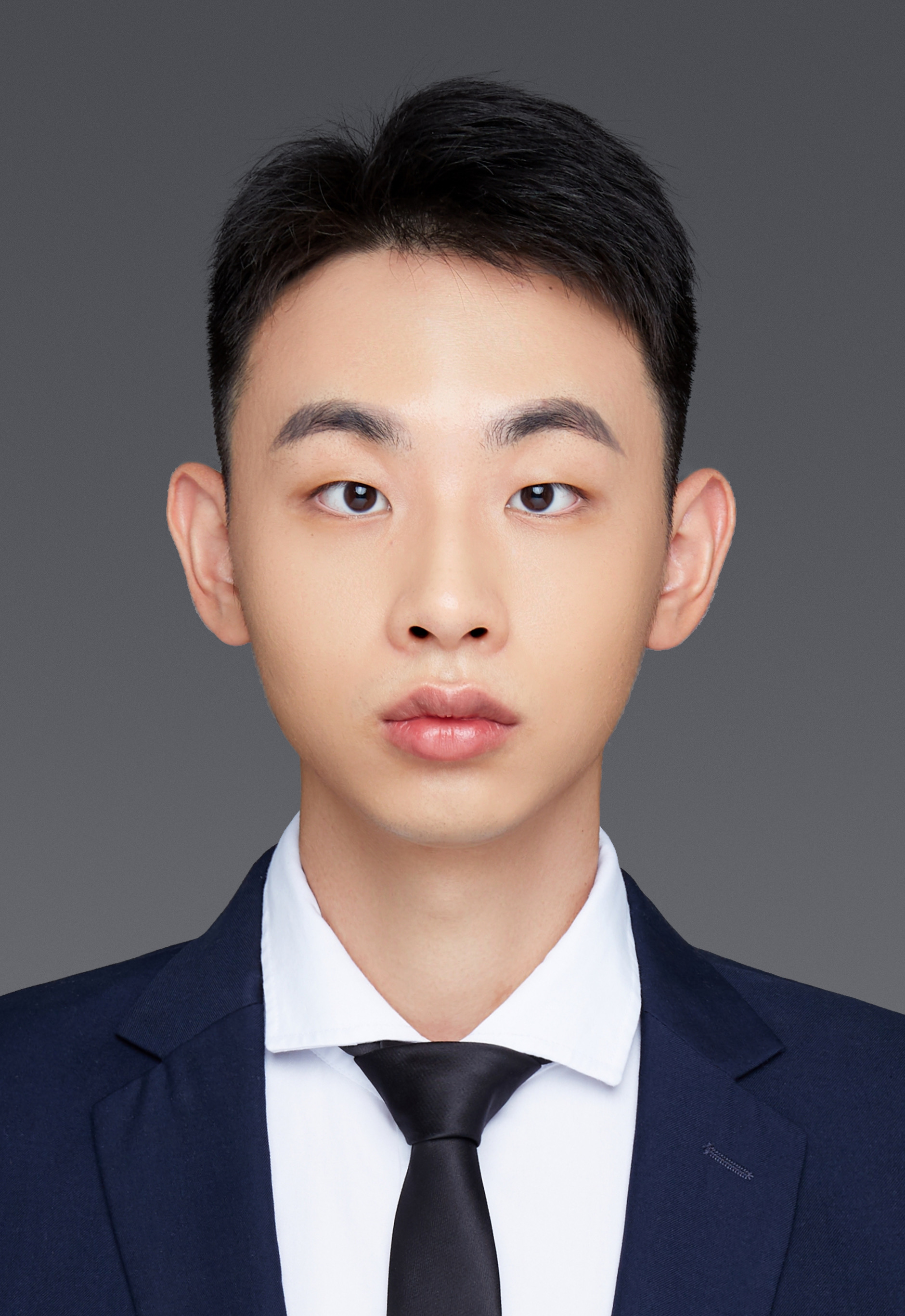}}]
	{Songjiang Yang} (Member, IEEE) is currently a postdoctoral researcher at the Pervasive Communication Research Center, Purple Mountain Laboratories, Nanjing, China. He received B.Eng. (First Class Hons.) and Ph.D. degrees in electrical and electronic engineering from the University of Sheffield, Sheffield, U.K., in 2017 and 2022 respectively. His current research interests include ray-tracing channel modeling, UAV  communications, and system performance evaluation. Dr. Yang was selected for the Outstanding Postdoctoral Fellow Program in Jiangsu Province and received the Best Paper Award from WCSP 2024.
\end{IEEEbiography}

\begin{IEEEbiography}[{\includegraphics[width=1in,height=1.25in,clip,keepaspectratio] {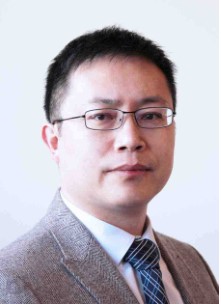}}]
	{Cheng-Xiang Wang} (Fellow, IEEE) received the B.Sc. and M.Eng. degrees in communication and information systems from Shandong University, China, in 1997 and 2000, respectively, and the Ph.D. degree in wireless communications from Aalborg University, Denmark, in 2004.
	
	He was a Research Assistant with the Hamburg University of Technology, Hamburg, Germany, from 2000 to 2001, a Visiting Researcher with Siemens AG Mobile Phones, Munich, Germany, in 2004, and a Research Fellow with the University of Agder, Grimstad, Norway, from 2001 to 2005. He was with Heriot-Watt University, Edinburgh, U.K., from 2005 to 2018, where he was promoted to a professor in 2011. He has been with Southeast University, Nanjing, China, as a professor since 2018, and he is now the Executive Dean of the School of Information Science and Engineering. He is also a professor with Pervasive Communication Research Center, Purple Mountain Laboratories, Nanjing, China. He has authored 4 books, 3 book chapters, and over 610 papers in refereed journals and conference proceedings, including 28 highly cited papers. He has also delivered 32 invited keynote speeches/talks and 21 tutorials in international conferences. His current research interests include wireless channel measurements and modeling, 6G wireless communication networks, and electromagnetic information theory.
	
	Dr. Wang is a Member of the Academia Europaea (The Academy of Europe), a Member of the European Academy of Sciences and Arts (EASA), a Fellow of the Royal Society of Edinburgh (FRSE), IEEE, and IET, an IEEE Communications Society Distinguished Lecturer in 2019 and 2020, a Highly-Cited Researcher recognized by Clarivate Analytics in 2017-2020. He is currently an Executive Editorial Committee Member of the IEEE TRANSACTIONS ON WIRELESS COMMUNICATIONS. He has served as an Editor for over sixteen international journals, including the IEEE TRANSACTIONS ON WIRELESS COMMUNICATIONS, from 2007 to 2009, the IEEE TRANSACTIONS ON VEHICULAR TECHNOLOGY, from 2011 to 2017, and the IEEE TRANSACTIONS ON COMMUNICATIONS, from 2015 to 2017. He was a Guest Editor of the IEEE JOURNAL ON SELECTED AREAS IN COMMUNICATIONS, the IEEE TRANSACTIONS ON BIG DATA, and the IEEE TRANSACTIONS ON COGNITIVE COMMUNICATIONS AND NETWORKING. He has served as a TPC Chair and General Chair for more than 30 international conferences. He received IEEE Neal Shepherd Memorial Best Propagation Paper Award in 2024. He also received 19 Best Paper Awards from international conferences.
	
\end{IEEEbiography}

\begin{IEEEbiography}[{\includegraphics[width=1in,height=1.25in,clip,keepaspectratio] {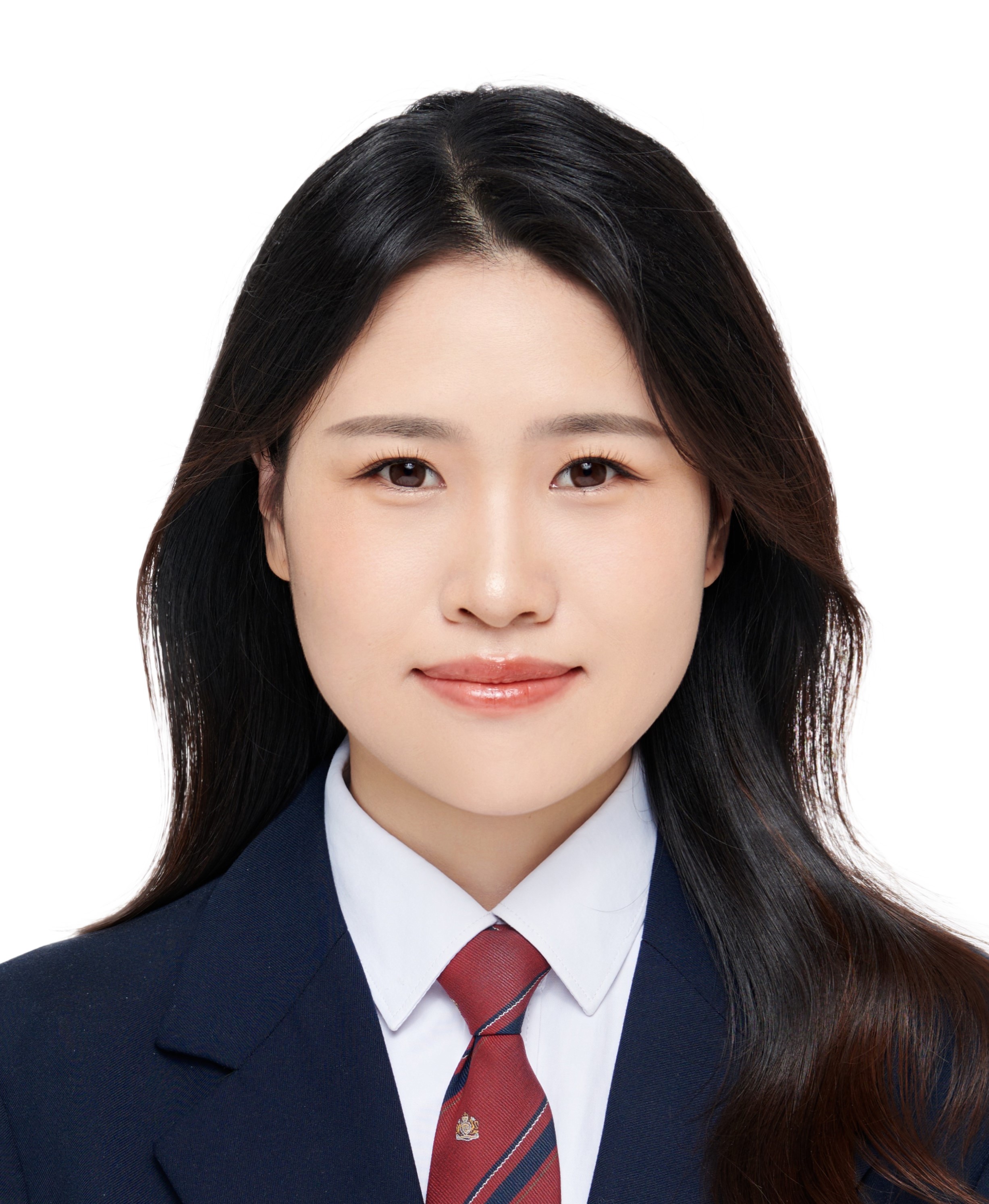}}]
	{Yinghua Wang} received the M.Sc. degree in Information and Communication Engineering from Harbin Institute of Technology, China, in 2017. She is currently a Wireless Channel Engineer in the Purple Mountain Laboratories, Nanjing, China. Her current research interests include 6G wireless channel modeling and channel simulators.
\end{IEEEbiography}

\begin{IEEEbiography}[{\includegraphics[width=1in,height=1.25in,clip,keepaspectratio] {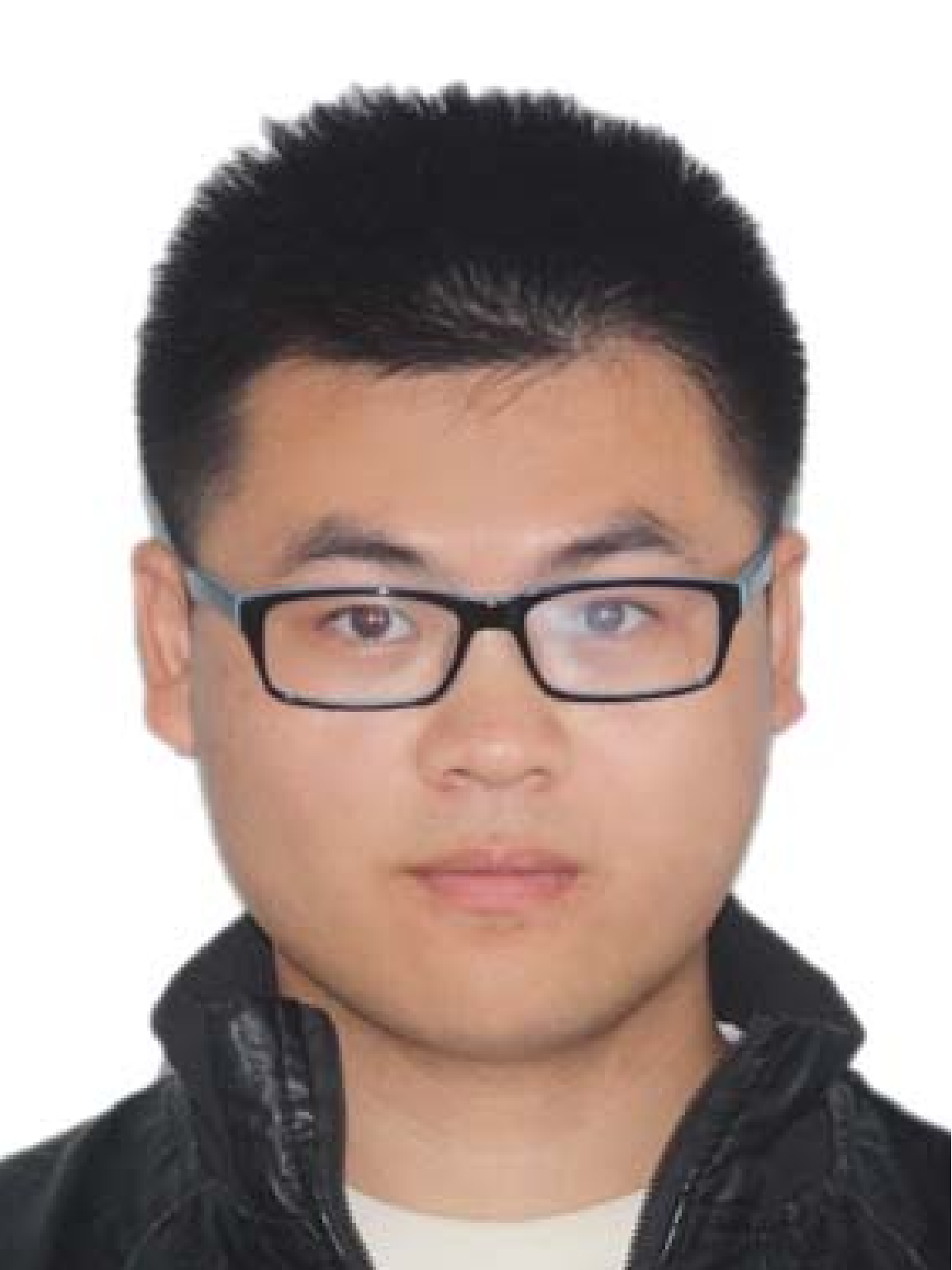}}]
	{Jie Huang} (Member, IEEE) received the B.E. degree in Information Engineering from Xidian University, China, in 2013, and the Ph.D. degree in Information and Communication Engineering from Shandong University, China, in 2018. From Oct. 2018 to Oct. 2020, he was a Postdoctoral Research Associate in the National Mobile Communications Research Laboratory, Southeast University, China, supported by the National Postdoctoral Program for Innovative Talents. From Jan. 2019 to Feb. 2020, he was a Postdoctoral Research Associate in Durham University, UK. Since Mar. 2019, he is a part-time researcher in Purple Mountain Laboratories, China. Since Nov. 2020, he is an Associate Professor in the National Mobile Communications Research Laboratory, Southeast University. He has authored and co-authored more than 100 papers in refereed journals and conference proceedings. He received the Best Paper Awards from WPMC 2016, WCSP 2020, WCSP 2021, and WCSP 2024. He has delivered over 10 tutorials in international conferences, including IEEE Globecom and IEEE ICC. His research interests include millimeter wave, massive MIMO, reconfigurable intelligent surface channel measurements and modeling, electromagnetic information theory, and 6G wireless communications.
\end{IEEEbiography}

\begin{IEEEbiography}[{\includegraphics[width=1in,height=1.25in,clip,keepaspectratio] {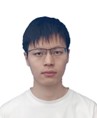}}]
	{Yuyang Zhou} received the B.Eng. degree in information science and engineering from the Southeast University, Nanjing, China, in 2022, where he is currently pursuing the Ph.D. degree. His research interests include wireless channel modeling and ray tracing channel modeling.
\end{IEEEbiography}

\begin{IEEEbiography}[{\includegraphics[width=1in,height=1.25in,clip,keepaspectratio] {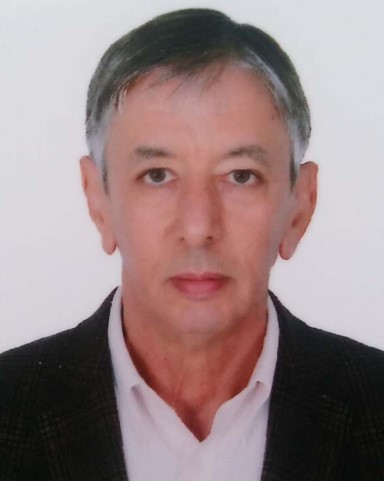}}]
	{El-Hadi M. Aggoune} (Life Senior Member, IEEE) received the M.S. and Ph.D. degrees in electrical engineering from the University of Washington (UW), Seattle, WA, USA. He served at several universities in the US and abroad at many academic ranks including Endowed Chair Professor. He is listed as Inventor in several patents, one of them assigned to the Boeing Company, USA. He is a Professional Engineer registered in the state of Washington. He co-authored papers in IEEE and other journals and conferences, and served on editorial boards and technical committees for many of them. He was the recipient of the IEEE Professor of the Year Award, UW. He was the director of a laboratory that received the Boeing Supplier Excellence Award. He is currently serving as Professor and Director of the AI and Sensing Technologies Research Center, University of Tabuk. His research interests include wireless communication, sensor networks, power systems, neurocomputing, and scientific visualization.
\end{IEEEbiography}

\end{document}